# Exotic decay channels are not the cause of the neutron lifetime anomaly

D. Dubbers,[1] H. Saul,[2] B. Märkisch,[2] T. Soldner,[3] and H. Abele[4]

[1] *Physikalisches Institut, Universität Heidelberg, Im Neuenheimer Feld 226, 69120 Heidelberg, Germany*

[2] *Physik-Department, Technische Universität München, James-Franck-Straße 1, 85748 Garching, Germany*

[3] *Institut Laue-Langevin, 71 avenue des Martyrs, CS 20156, 38042 Grenoble Cedex 9, France*

[4] *Atominstitut, Technische Universität Wien, Stadionallee 2, 1020 Wien, Austria*

*Abstract*. Since long neutron lifetimes measured with a beam of cold neutrons are significantly different from lifetimes measured with ultracold neutrons bottled in a trap. It is often speculated that this "neutron anomaly" is due to an exotic dark neutron decay channel of unknown origin. We show that this explanation of the neutron anomaly can be excluded with a high level of confidence when use is made of our new result for the neutron decay $\beta$ asymmetry. Furthermore, data from neutron decay now compare well with $Ft$-data derived from nuclear $\beta$ decays.

**Introduction**

Neutron $\beta$ decay plays a key role in several fields of physics and astrophysics [1], [2], [3], [4]. On one hand, all semileptonic processes in nature, which involve both first-generation quarks and leptons, require neutron decay data for the calculation of their cross sections or rates. On the other hand, neutron data are increasingly used for sensitive searches of new physics beyond the standard model (SM). Rigorous bounds on parameters beyond the SM can be derived from low-energy processes like neutron, pion, or nuclear weak decays, and from high-energy processes like $d\bar{u} \to e\bar{\nu}_e$ in $p\bar{p}$ reactions at the LHC. On the quark level, the latter reaction has the same Feynman diagram as neutron decay $d \to u e\bar{\nu}_e$. With effective field theory [5], these high- and low-energy processes can be linked and data compared with each other. In many cases the low-energy data lead to better constraints, in particular for processes beyond the SM involving left-handed (SM-)neutrinos. In contrast, high-energy data from LHC give better limits on processes involving right-handed neutrinos, see [6], [7], [8], and references therein.

Over the years the precision of neutron decay data has seen considerable progress. In the past three decades, errors of the neutron lifetime have diminished by a factor of ten, and errors of the $\beta$ decay asymmetry by a factor of twenty, see the previous editions of the Particle Data Group's reviews (PDG) [9]. At the same time, these data have become more reliable: the corrections required to obtain the neutron lifetime from the raw data have dropped from hundreds of seconds to one quarter of a second, and the leading corrections to the $\beta$ asymmetry diminished more than tenfold, as found in the corresponding literature.

So everything seems to proceed well, but there is a rather longstanding problem. The neutron lifetime can be measured with two different methods, and since many years, the lifetimes derived from these differ significantly [9], [10], [11], [12]. Most lifetime experiments nowadays use the decay of ultracold neutrons (UCN) stored in a trap, as pioneered by W. Mampe et al. [13]. In these "bottle" experiments, the exponential decrease of the number of stored UCN is registered. In the "beam" experiments, a beam of cold neurons is used, and the decay products emitted from a well-defined beam volume are counted. Today, the average bottle lifetime, derived from eight measurements on five different instruments, is by four standard deviations shorter than the average beam lifetime, the latter being obtained from two runs of one instrument.

It is frequently speculated that this "neutron anomaly" might be due to an exotic neutron decay into a dark fermion. Such a decay channel would be visible in the total decay rate of the bottle experiments, but not in the beam experiments. Various possible dark decay channels have been discussed in very recent papers, of which we give an incomplete list: Investigated were exotic decays that are completely dark, or with the dark fermion accompanied either by visible particles such as $\gamma$ or $e^+e^-$ [14], or by invisible $\nu\bar{\nu}$ pairs [15] or dark photons [16]. The disappearance of neutrons via neutron mirror-neutron oscillations was proposed in [17], [18], and the role of neutron-antineutron oscillations in dark neutron decay investigated in [19]. In Ref. [20] it is pointed out that a Fierz term of size $b = 1.44\%$ would enhance the branching ratio of dark decays, allowed by existing neutron data, to the level required to explain the neutron lifetime anomaly. This in turn, however, leads to some tension with the experimental limit on $b$ from [21]. The detection of neutron dark decays via nuclear decays was discussed in [22], [23], and detection by electro-disintegration of the deuteron in [24]. According to [25], such dark decays could also solve problems in the small-structure formation in cosmology.

However, neutron dark decays would lead to problems with observed neutron star masses [26], [27], [28]. Dark neutron decays that are accompanied by $\gamma$ [29] or $e^+e^-$ emission [30], [31] were experimentally excluded as cause of the neutron anomaly for most of the relevant energy range. It would be desirable to verify or to exclude dark neutron decays on a more general basis. Since several years, the neutron anomaly has also reached the popular science sector, see [32], [33], and others. Due to this, the public is aware of the neutron anomaly, but not of the strong progress made in neutron decay.

In the SM, the neutron lifetime $\tau_\beta$ for the decay $n \to pe^-\bar{V}_e$ and its axial-vector coupling constant $g_A$ are linked to each other in a well-known way. A recent letter [34] suggested to use this link to test the hypothesis of dark neutron decay. However, the lifetime and $g_A$ were not known with sufficient precision for this purpose. Therefore, the authors made an educated guess on "favored" values for lifetime and $g_A$ that would satisfy this link and provide a bound on the branching ratio for dark neutron decays.

In the present letter we show that we can now test the hypothesis of a dark branch in neutron decay, like in Ref. [34], though slightly modified, and based not on favored values but on measured data that include all experimental results on neutron decay. This has become possible by including new results on the $\beta$ decay asymmetry not yet listed in PDG-2018. In the following, we first explain the method in some detail, and then discuss the new data base and its consequences for the neutron anomaly.

**The method**

In the *beam* experiments on the neutron lifetime, cold neutrons in a beam are absorbed in a neutron detector within milliseconds after they have entered the decay volume. Therefore, the number $N$ of neutrons in this fiducial volume does not depend significantly on the value of the neutron lifetime. The rate of electron or proton emission from the decay volume then is $n_\beta = N/\tau_\beta$, with the lifetime $\tau_\beta$ for ordinary neutron decay $n \to pe^-\bar{V}_e$, while possible dark decays go undetected. The decay rate measured in a beam experiment therefore equals the true partial $\beta$ decay rate,

$$\tau_{\text{beam}}^{-1} = \tau_\beta^{-1}. \tag{1}$$

In the *bottle* experiments, the UCN remaining in the trap decay via both channels, allowed and exotic, as $N(t) = N(0)\exp(-t/\tau_{\text{bottle}})$, where

$$\tau_{\text{bottle}}^{-1} = \tau_\beta^{-1} + \tau_X^{-1}, \tag{2}$$

with the partial decay rate $\tau_X^{-1}$ into unknown channels $X$, and we set the overall neutron lifetime $\tau_n \equiv \tau_{\text{bottle}}$. Hence $\tau_{\text{bottle}} < \tau_{\text{beam}}$ as it is observed. The lifetime $\tau_{\text{bottle}}$ is obtained by measuring $N(t)$ for several different storage intervals $t$. (We assume that other more mundane losses from the UCN trap are corrected for.)

With the measured lifetimes $\tau_{\text{bottle}} \approx 880(1)\,\text{s}$ and $\tau_{\text{beam}} \approx 888(2)\,\text{s}$ we obtain the frequently quoted branching ratios for partial $\beta$ decay $\text{BR}_\beta \equiv \tau_\beta^{-1}/\tau_n^{-1}$, or

$$\text{BR}_\beta = \tau_{\text{bottle}}/\tau_{\text{beam}} = 99.0(0.2)\%, \tag{3}$$

and for decay into dark channels $X$,

$$\text{BR}_X = 1 - \text{BR}_\beta = 1.0(0.2)\%. \tag{4}$$

The above-mentioned link between $\tau_\beta$ and $g_A$ is given by the so-called SM master formula [34], which allows calculating the neutron lifetime expected in the SM,

$$\tau_\beta^\lambda = \frac{4908.6(1.9)\,\text{s}}{|V_{ud}|^2(1+3\lambda^2)}, \tag{5}$$

from a given value of the ratio $\lambda = g_A/g_V$ of the neutron weak axial-vector to vector couplings. Therein, the number in the denominator is a combination of the fundamental constants $c$, $\hbar$, and $m_e$. Ref. [34] makes use of this link by inserting the CKM matrix element $V_{ud}$ as derived from nuclear superallowed $\beta$ decays. The leading error of $V_{ud}$ comes from the universal radiative correction $\Delta_R^V$, which must then be eliminated because the right-hand side of Eq. (5) is independent of $\Delta_R^V$. Instead, we use Eq. (9) of Ref. [35]

$$\tau_\beta^\lambda = \frac{2}{\ln 2}\frac{\overline{Ft}_{0^+\to 0^+}}{f(1+\delta_R')(1+3\lambda^2)} = \frac{5172.3(1.1)\,\text{s}}{1+3\lambda^2}. \tag{6}$$

This result is consistent with that of Ref. [34], but the ingredients on the right-hand side of Eq. (6) are independent of $\Delta_R^V$. In this equation, the average of nuclear superallowed $Ft$ values $\overline{Ft}_{0^+ \to 0^+} = 3072.27(0.62)\,\text{s}$ is taken from Ref. [36], for more details see the discussion on $Ft$ in our last section below.

Hence, if the dark-channel hypothesis is right, then the SM-lifetime $\tau_\beta^\lambda$ calculated from Eq. (6) should coincide with $\tau_{\text{beam}}$ and not with the shorter $\tau_{\text{bottle}}$. To find out we need to know precisely $\lambda = g_A / g_V$.

- The value of $\lambda$ can in principle be derived from lattice theory, but presently only with a precision of 1% [37], which is by far not sufficient for our purpose.

- Experimentally, the value of $\lambda$ is derived from neutron decay correlation coefficients, which in the SM all depend only on $\lambda$. The coefficients most sensitive to $\lambda$ are the β decay asymmetry $A$ and the electron-antineutrino correlation $a$,

$$A = -2\frac{\lambda(\lambda+1)}{1+3\lambda^2}, \quad a = \frac{1-\lambda^2}{1+3\lambda^2}, \tag{7}$$

because both coefficients respond to the deviation of $\lambda$ from $-1$.

- The PDG-2018 average derived from these equations is $\lambda = -1.2724(23)$. Inserted into Eq. (5), this gives $\tau_\beta^\lambda = 883(2)\,\text{s}$, almost half way between $\tau_{\text{beam}} = 888.0(2.0)\,\text{s}$ and the (updated) $\tau_{\text{bottle}} = 879.4(0.6)\,\text{s}$, so this does not help to decide between the two.

- For their choice of data, the authors of Ref. [34] took into consideration only the bottle lifetimes $\tau_{\text{bottle}}$, and only the data $g_A$ from year 2002 on, and required that they are compatible with Eq. (6), to arrive at their choices $\tau_{\text{favored}} = 879.4(0.6)\,\text{s}$ and $\lambda_{\text{favored}} = -1.2755(11)$. With these values they obtained an upper bound for the dark branching ratio BR$_X$ < 0.27% (95% C.L.), while BR$_X$ = 1.0 (0.2)% would be needed to explain the neutron anomaly.

**The data base**

The past months have seen a flurry of new neutron decay data, which we added to the list of PDG-2018: three measurements of $\tau_{\text{bottle}}$, two measurements of $A$, and one of $a$. (For references to the previous data, see PDG-2018).

- The three new bottle lifetimes [38], [39], [40] confirm earlier bottle measurements; the corresponding preprints are already cited in Ref. [34]. The new data only slightly change the bottle lifetime average, from $\tau_{bottle} = 879.6(0.7)$ s in PDG-2018, where the error is increased by a scale factor $S = 1.2$, to $\tau_{bottle} = 879.4(0.6)$ s in our update of PDG-2018 (identical to $\tau_{favored}$), with the scale factor increased to $S = 1.5$, due to the scatter in the new data.

- The new electron-antineutrino value from *a*SPECT [41] has a four times lower error than previous *a*-values, but is preliminary and therefore not used here, but its inclusion would not significantly change the conclusion of our analysis.

- The new $\beta$ asymmetry measurements are crucial for our discussion. Fig. 1 shows the asymmetry values No. 1 to 5 that entered the PDG-2018 average, and the new data No. 6 and No. 7.

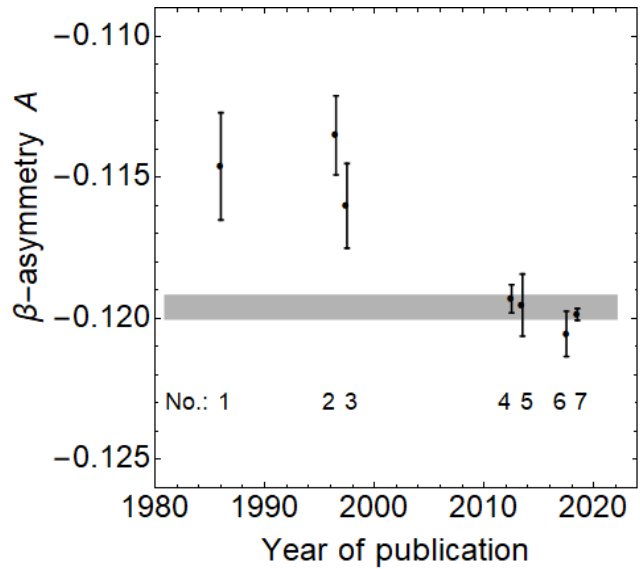

Fig. 1. To the $\beta$ asymmetry data that entered the PDG-2018 average (No. 1 to 5), we add recent results from UCNA (No. 6) and from PERKEO III (No. 7). The gray-shaded horizontal line indicates the weighted mean of the data and its one sigma error.

The data points No. 4 and No. 7 are from the cold-beam instruments PERKEO II [35] and III [42], respectively. PERKEO III at ILL uses a cold beam of polarized neutrons, pulsed with a duty cycle of 1:7, such that a free "cloud" of neutrons of high density is moving along the beam axis through the instrument. The decay electrons emitted from this cloud are projected magnetically onto energy-sensitive plastic scintillation detectors without meeting any material obstacle and without any edge effects. From the peak electron rate $n_\beta \approx > 1000$ s$^{-1}$ we conclude that the number of cold polarized neutrons in each such pulse is $N = n_\beta \tau_\beta \approx 10^6$, in accordance with Monte Carlo simulations of the setup.

The data points No. 5 and No. 6 are from the bottle instrument UCNA [43], [44]. UCNA at LANSCE has typically 4000 UCNs stored in a cylindrical bottle with material walls, with two thin windows for the magnetically guided electrons to leave the bottle. Thin $\Delta E$ gas detectors in front of the plastic $E$-scintillators are used to reduce background. A continuous electron rate of $n_\beta = 25$ s$^{-1}$ is obtained, at a very low background of 0.025 s$^{-1}$ [43]. Four UCN populations with different histories are encountered, which are carefully disentangled by separate measurements. Otherwise, UCNA and PERKEO III have precisely known neutron polarizations of 99.60(20)% and 99.10(06)%, respectively, and both use blinded analysis.

The weighted mean of all data in Fig. 1 is $A = -0.1196(4)$, as compared to the PDG-2018 average $A = -0.1184(10)$, where in both cases scale factor is $S = 2.4$. There is a certain dilemma concerning the scale factor $S$. To curb the influence of earlier data of lower quality, PDG excludes from the calculation of the scale factor $S$ all data points $A_i$ whose error $\sigma_i$ is larger than a critical value $\sigma_0 \equiv 3 \times \sqrt{N}\sigma$ (for $N$ data with average unscaled error $\sigma$), without excluding the data points from the weighted mean $A$ and its error $\sigma$. We find that the data points No. 1 to 3 have errors near or above the critical value $\sigma_0 = 0.00143$, namely, $\sigma_1 = 0.0019$, $\sigma_2 = 0.0014$, and $\sigma_3 = 0.0015$. Exclusion of all three data from the calculation of $S$ leads to $S = 0.86$, which, reset to $S = 1$, would considerably diminish the error of the average $A$. But No. 2 is a border case, and exclusion of only No. 1 and No. 3 leads to the same $S = 2.4$ as before. So we stay conservative and do not reduce $S$ and use all data for the evaluation of $A$.

PDG-2018 had arrived at an average $\lambda = -1.2724(23)$ with $S = 2.2$. When we update their list with the $\lambda$-values from the new $A$ measurements, this gives $\lambda = -1.2756(9)$ with $S = 2.4$. This value is indeed very close to $\lambda_{\text{favored}} = -1.2755(11)$ of Ref. [34], but is based on a full set of measured neutron data. (Outlook: Should someday PDG decide to drop the data points No. 1 to 3, then $\lambda = -1.2762(5)$, with a considerably smaller error).

**Consequences for the dark-decay hypothesis**

Inserted into Eq. (6), our $\lambda = -1.2756(9)$ gives $\tau_\beta^\lambda = 879.4(0.8)\,\mathrm{s}$, which coincides with the (updated) $\tau_{\mathrm{bottle}} = 879.4(0.6)\,\mathrm{s}$, see Fig. 2, and not with $\tau_{\mathrm{beam}} = 888.0(2.0)\,\mathrm{s}$, as one would expect if the neutron anomaly was due to an exotic branch. We emphasize that the results from PERKEO III, UCNA, and UCNτ that enter Fig. 2 are derived from blinded data. This leaves not much room for a dark channel in neutron decay.

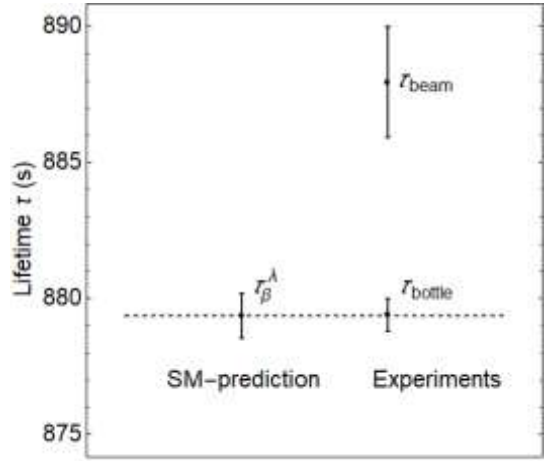

Fig. 2. The standard model expectation for the neutron lifetime $\tau_\beta^\lambda$ from Eq. (6) coincides with the measured bottle lifetime, and not with the beam lifetime. This finding excludes a dark branch as cause of the neutron anomaly. The dashed line through $\tau_\beta^\lambda$ is inserted to guide the eye.

Like the other publications on the neutron anomaly, we assume that the parameters entering the analysis, in particular the nuclear $Ft$-values whose average is used in Eq. (6), are not affected by the exotic process in question. But even if they were, it would require some fine-tuning to shift $\tau_\beta^\lambda$ to $\tau_{\mathrm{beam}}$. In addition for most nuclei, nuclear dark decays are forbidden due to energy constraints, see [22] and [23]. To add a very unlikely possibility: should the difference between $\tau_\beta^\lambda$ and $\tau_{\mathrm{beam}}$ be merely a statistical outlier of probability $6 \times 10^{-5}$, then this probability is not much higher than the probability of $4 \times 10^{-5}$ that the neutron anomaly itself is due to a statistical error.

We can also calculate a new bound on BR$_X$. Like in Ref. [34], we calculate one-sided bounds with 95% C.L. In addition, we use truncated distributions to account for the upper constraint BR ≤ 100%, which increases all bounds slightly, for instance the guessed bound in Ref. [34] from 0.27%, to 0.32%. We use the updated values for $\tau_{\text{bottle}}$ and $\lambda$ in Eq. (12) of Ref. [34] and find the new bound $BR_X < 0.28\%$. This bound is 3.3 times better than the bound $BR_X < 0.92\%$ derived from the data of PDG-2018 alone, which latter is still compatible with the value $BR_X = 1.0(0.2)\%$ from Eq. (4). When we discard the three $\lambda$ values from the past century, as was done in Ref. [34], our bound drops to $BR_X < 0.14\%$. We conclude that the discussions of dark neutron decays (interesting as they are) should no longer be pursued in the context of the neutron anomaly.

**The *Ft* value for neutron decay**

We use the occasion to point out that, with the new neutron decay data cited in this article, the neutron-derived *Ft*-value becomes competitive with the *Ft*-values of superallowed nuclear $0^+ \to 0^+$ $\beta$ decays [36], which latter is

$$Ft_{0^+ \to 0^+} \equiv f t_{0^+ \to 0^+} (1+\delta'_R)(1+\delta_{NS}-\delta_C). \tag{8}$$

with nuclear half-lives $t$ and phase space factors $f$. In this equation, $\delta'_R$ and $\delta_{NS}$ are the nuclear transition-dependent radiative corrections, and $\delta_C$ is the isospin correction. $\delta'_R$ is a function only of nuclear charge $Z$ and $\beta$ energy $E$, independent of nuclear structure, and typically close to 1.5%; $\delta_{NS}$ and $\delta_C$ are in most cases a fraction of 1%, see Table X in [36].

Under CVC, Eq. (8) holds also for the vector part of neutron decay, with an additional spin factor ½. For the neutron, nuclear-structure dependent corrections are absent, $\delta_{NS} = \delta_C = 0$. The neutron's branching ratio for Fermi transitions equals $1/(1+3\lambda^2)$, and we need $\lambda$ as additional parameter (likewise, for $\beta$ transitions to different nuclear levels, separately measured branching ratios are needed to obtain $Ft_{0+\to 0+}$). The vector part of the neutron *Ft*-value is therefore

$$Ft_{nV} \equiv f t_{nV}(1+\delta'_R) = \tfrac{1}{2}\ln 2 \, f\tau_n \, (1+3\lambda^2)(1+\delta'_R), \tag{9}$$

with $f = 1.6887(2)$ and $\delta'_R = 0.014902(2)$ from Sect. 6.2 of [45], both known with high-precision, and with the measured neutron lifetime $\tau_n$. When we replace, under CVC, $Ft_{nV}$ in Eq. (9) with twice the average $2\overline{Ft}_{0^+\to 0^+}$ over the nuclei, we are back to Eq. (6). There is no dependence on $\Delta^V_R$, which is fortunate because a recent calculation suggests [46], [47] that the last word on its value may not yet have been spoken.

Fig. 3 shows the nuclear $Ft$-values in dependence of $Z$ of the daughter nuclei, as taken from Ref. [36]. The horizontal gray-shaded band and its width indicate the average of the nuclear values and its error, $\overline{Ft}_{0^+\to 0^+} = 3072.27(0.62)\,\text{s}$. At $Z=1$ we added the neutron result from Eq. (9), $Ft_{nV} = 3073.6(3.9)\,\text{s}$. This result is based on all neutron data for lifetime $\tau_n$ and $\lambda$. Our interpretation of Fig. 3 is that neutron decay data nowadays compare well with the data derived from nuclear decays. If someday the old $A$ values No. 1-3 are excluded from the calculation of the scale factor $S$ (or are excluded altogether), then the error of the neutron's $Ft_{nV}$ will be reduced further.

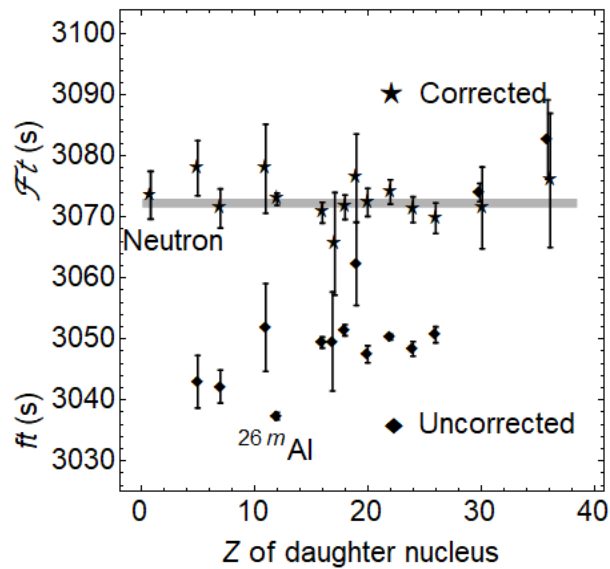

Fig. 3. The neutron $Ft$-value at $Z=1$, derived from all available neutron data (update of PDG-2018), has no nuclear corrections and compares well with the individual $Ft$-values for superallowed nuclear $\beta$ transitions, as taken from [36]. The horizontal gray-shaded band and its width indicate the average of the nuclear $Ft_{0^+\to 0^+}$-values and its error.

**Conclusion**

It is often speculated that the neutron decay anomaly may be due to dark neutron decay channels. Our analysis, based on all neutron decay data, excludes such an explanation, cf. Fig. 2, and lowers the bound on the dark branching ratio from $BR_X < 0.92\%$ (95% C.L.), based on the data of PDG-2018, to $BR_X < 0.28\%$, based on our update of PDG-2018. We also show, Fig. 3, that neutron decay data nowadays compare well with $Ft$-data derived from nuclear $\beta$ decays.

**Acknowledgment**

This work was supported by the DFG Priority Programme SPP 1491 and the Austrian Science Fund (FWF).